\documentstyle[psfig]{mn}

\newif\ifAMStwofonts
\ifoldfss
  \ifCUPmtlplainloaded \else
    \NewTextAlphabet{textbfit} {cmbxti10} {}
    \NewTextAlphabet{textbfss} {cmssbx10} {}
    \NewMathAlphabet{mathbfit} {cmbxti10} {} % for math mode
    \NewMathAlphabet{mathbfss} {cmssbx10} {} %  "   "    "
  \fi
  \ifAMStwofonts
    \ifCUPmtlplainloaded \else
      \NewSymbolFont{upmath} {eurm10}
      \NewSymbolFont{AMSa} {msam10}
      \NewMathSymbol{\upi}     {0}{upmath}{19}
      \NewMathSymbol{\umu}     {0}{upmath}{16}
      \NewMathSymbol{\upartial}{0}{upmath}{40}
      \NewMathSymbol{\leqslant}{3}{AMSa}{36}
      \NewMathSymbol{\geqslant}{3}{AMSa}{3E}

       \let\le=\leqslant
       \let\ge=\geqslant
    \fi
  \fi
\fi % End of OFSS

\ifnfssone
  \newmathalphabet{\mathit}
  \addtoversion{normal}{\mathit}{cmr}{m}{it}
  \addtoversion{bold}{\mathit}{cmr}{bx}{it}
  \newmathalphabet{\mathbfit} % math mode version of \textbfit{..}
  \addtoversion{normal}{\mathbfit}{cmr}{bx}{it}
  \addtoversion{bold}{\mathbfit}{cmr}{bx}{it}
  \newmathalphabet{\mathbfss} % math mode version of \textbfss{..}
  \addtoversion{normal}{\mathbfss}{cmss}{bx}{n}
  \addtoversion{bold}{\mathbfss}{cmss}{bx}{n}
  \ifAMStwofonts
    \ifCUPmtlplainloaded \else
      \UseAMStwoboldmath
      \makeatletter
      \new@mathgroup\upmath@group
      \define@mathgroup\mv@normal\upmath@group{eur}{m}{n}
      \define@mathgroup\mv@bold\upmath@group{eur}{b}{n}
      \edef\UPM{\hexnumber\upmath@group}
      \new@mathgroup\amsa@group
      \define@mathgroup\mv@normal\amsa@group{msa}{m}{n}
      \define@mathgroup\mv@bold\amsa@group{msa}{m}{n}
      \edef\AMSa{\hexnumber\amsa@group}
      \makeatother
      \mathchardef\upi="0\UPM19
      \mathchardef\umu="0\UPM16
      \mathchardef\upartial="0\UPM40
      \mathchardef\leqslant="3\AMSa36
      \mathchardef\geqslant="3\AMSa3E

       \let\le=\leqslant
       \let\ge=\geqslant
    \fi
  \fi
\fi % End of NFSS release 1

\ifnfsstwo
  \DeclareMathAlphabet{\mathbfit}{OT1}{cmr}{bx}{it}
  \SetMathAlphabet\mathbfit{bold}{OT1}{cmr}{bx}{it}
  \DeclareMathAlphabet{\mathbfss}{OT1}{cmss}{bx}{n}
  \SetMathAlphabet\mathbfss{bold}{OT1}{cmss}{bx}{n}
  \ifAMStwofonts
    \ifCUPmtlplainloaded \else
      \DeclareSymbolFont{UPM}{U}{eur}{m}{n}
      \SetSymbolFont{UPM}{bold}{U}{eur}{b}{n}
      \DeclareSymbolFont{AMSa}{U}{msa}{m}{n}
      \DeclareMathSymbol{\upi}{0}{UPM}{"19}
      \DeclareMathSymbol{\umu}{0}{UPM}{"16}
      \DeclareMathSymbol{\upartial}{0}{UPM}{"40}
      \DeclareMathSymbol{\leqslant}{3}{AMSa}{"36}
      \DeclareMathSymbol{\geqslant}{3}{AMSa}{"3E}

       \let\le=\leqslant
       \let\ge=\geqslant
    \fi
  \fi
\fi % End of NFSS release 2

\ifCUPmtlplainloaded \else
  \ifAMStwofonts \else % If no AMS fonts
    \def\upi{\pi}
    \def\umu{\mu}
    \def\upartial{\partial}
  \fi
\fi

\title{A deep $UBVRI$ CCD photometric study of the open clusters Tr 1 and Be 11}
\author[R. K. S. Yadav and Ram Sagar]
       {R. K. S. Yadav$^{1}$\thanks{E-mail: rkant@upso.ernet.in} and Ram Sagar$^{1,2}$\thanks{E-mail: sagar@upso.ernet.in}\\
        $^{1}$State Observatory, Manora Peak Nainital 263129, India\\
        $^{2}$Indian Institute of Astrophysics, Bangalore 560034, India}
\date{Accepted ---------.
      Received ---------;
      }

\pagerange{\pageref{firstpage}--\pageref{lastpage}}
\pubyear{2002}
\begin{document}
\maketitle
\label{firstpage}
\begin{abstract}
We present deep $UBVRI$ CCD photometry for the young open star clusters Tr 1 
and Be 11. The CCD data for Be 11 is obtained for the first time. The sample 
consists of $\sim$ 1500 stars reaching down to $V$ $\sim$ 21 mag. Analysis 
of the radial distribution of stellar surface density indicates that radius 
values for Tr 1 and Be 11 are 2.3 and 1.5 pc respectively. The interstellar 
extinction across the face of the imaged clusters region seems to be 
non-uniform with a mean value of $E(B-V)$ = 0.60$\pm$0.05 and 0.95$\pm$0.05 
mag for Tr 1 and Be 11 respectively. A random positional variation of $E(B-V)$ 
is present in both the clusters. In the cluster Be 11, the reason of random 
positional variation may be apparent association of the HII region (S 213). The 
2MASS $JHK$ data in combination with the optical data in the cluster Be 11 
yields $E(J-K)$ = 0.40$\pm$0.20 mag and $E(V-K)$ = 2.20$\pm$0.20 mag. 
Colour excess diagrams indicate a normal interstellar extinction law in the 
direction of cluster Be 11.\\
The distances of Tr 1 and Be 11 are estimated as 2.6$\pm$0.10 and 2.2$\pm$0.10 
Kpc respectively, while the theoretical stellar evolutionary isochrones
fitted to the bright cluster members indicate that the cluster Tr 1 and 
Be 11 are 40$\pm$10 and 110$\pm$10 Myr old. The mass functions corrected for 
both field star contamination and data incompleteness are derived for both the clusters.
The slopes $1.50\pm0.40$ and $1.22\pm0.24$ for Tr 1 and Be 11 respectively are  
in agreement with the Salpeter's value. Observed mass segregations in both 
clusters may be due to the result of dynamical evolutions or 
imprint of star formation processes or both.\\

\end{abstract}


\section{Introduction}

Young open star clusters in a galaxy provide valuable information
about star formation processes and are key objects for the
galactic structure and evolution. For such studies, a knowledge of cluster's
parameters like distance, age, reddening and stellar content is
required which can be derived from the colour-magnitude (CM) and colour-colour 
(CC) diagrams of star clusters. In addition to this, the distribution of 
stellar masses at the time of cluster formation is of fundamental 
importance to analysis related to evolution of galaxies. The initial mass 
function (IMF) also plays an important role in understanding the early 
dynamical evolution of star clusters, because it is a fossil record of the 
very complex process of star formation and provides an important link between 
the easily observable population of luminous stars in a stellar system and the 
fainter, but dynamically more important, low mass stars. Another related 
problem is the mass segregation in star clusters in which massive stars are 
more concentrated towards the cluster center compared to low mass stars. It is 
not clear whether the mass segregation observed in several open clusters is due
 to dynamical evolution or an imprint of star formation processes itself 
(cf. Sagar et al. 1988; Sagar 2001 and references therein). Thus one can say 
that the young open star clusters are the laboratories in a galaxy for 
providing answers to many current questions of astrophysics.\\

\begin{table*}
 \centering
 \begin{minipage}{140mm}
 \caption{General information about the clusters under study, taken from 
Mermilliod (1995)}
\begin{tabular}{|c|ccccccccc|}
\hline
Cluster&IAU&OCL&l&b&Trumpler&Radius&Distance&$E(B-V)$&log(age)\\
&&&(deg)&(deg)&class&(arcmin)&(Kpc)&(mag)&(yrs)\\
\hline
Trumpler 1&C0132+610&328&128.22&-1.14&II 2p&1.5&2.6&0.58&7.5\\
Berkeley 11&C0417+448&404&157.08&-3.65&II 2m&2.5&2.2&0.95&7.7\\
\hline
\end{tabular}
\end{minipage}
\end{table*}

       In the light of above discussions, we performed multicolour deep CCD 
stellar photometry in two young open star clusters namely Trumpler 1 (Tr 1) 
and Berkeley 11 (Be 11). The CCD $UBVRI$ observations of Be 11 are presented 
for the first time. The relevant prior informations (taken from Mermilliod 
(1995)) of these clusters are given in Table 1. These clusters are relatively 
compact objects with angular radii less than $3^{\prime}$. Previous studies,
 observations and data reductions are described in the next sections. The
interstellar extinction, other photometric results, luminosity function,
mass function and mass segregation are described in
the subsequent sections. Making use of $JHK$ data with optical data, extinction
law has also been studied in Be 11.

\section[]{Previous studies}
{\bf Trumpler 1}: It is an extremely concentrated galactic open star
cluster in Cassiopeia. It lies at the outer edge of the Perseus spiral arm.
Oja (1966) carried out the proper motion study. $UBV$ photoelectric photometry
for 43 bright stars was presented by Joshi \& Sagar (1977) while Phelps \& Janes
 (1994) published $UBV$ CCD photometry ($V$$\sim$18 mag). McCuskey \& Houk 
(1964) studied this cluster photographically in $UBV$ system
while Steppe (1974) has presented three colour RGU photographic photometry.
 All these studies indicate that reddening across the cluster is uniform with 
$E(B-V)$ = 0.61 mag, distance estimate is 2630 pc and age seems to be $\sim$ 
27 Myr.\\

\noindent {\bf Berkeley 11}: It is a distant neglected compact young open cluster, 
apparently associated with the faint HII region S 213. Only $UBV$ 
photoelectric photometry of 24 bright stars has been done by Jackson et al. 
(1980). On the basis of this study, they found that this cluster has members 
earliest photometric type $\sim$ b4 and is thus an extreme Population I 
object. They also found reddening $E(B-V)$ = 0.95$\pm$0.06 mag, distance d = 
2.2$\pm$ 0.2 Kpc and age of the cluster as 3$\times$10$^{7}$ yr.\\
\section{Observational data}
The optical and near-IR $JHK$ data used in the present study are described
in the following subsections.

\subsection{Optical observations and data reductions}

\begin{figure*}
\hspace*{1.0cm}\psfig{file=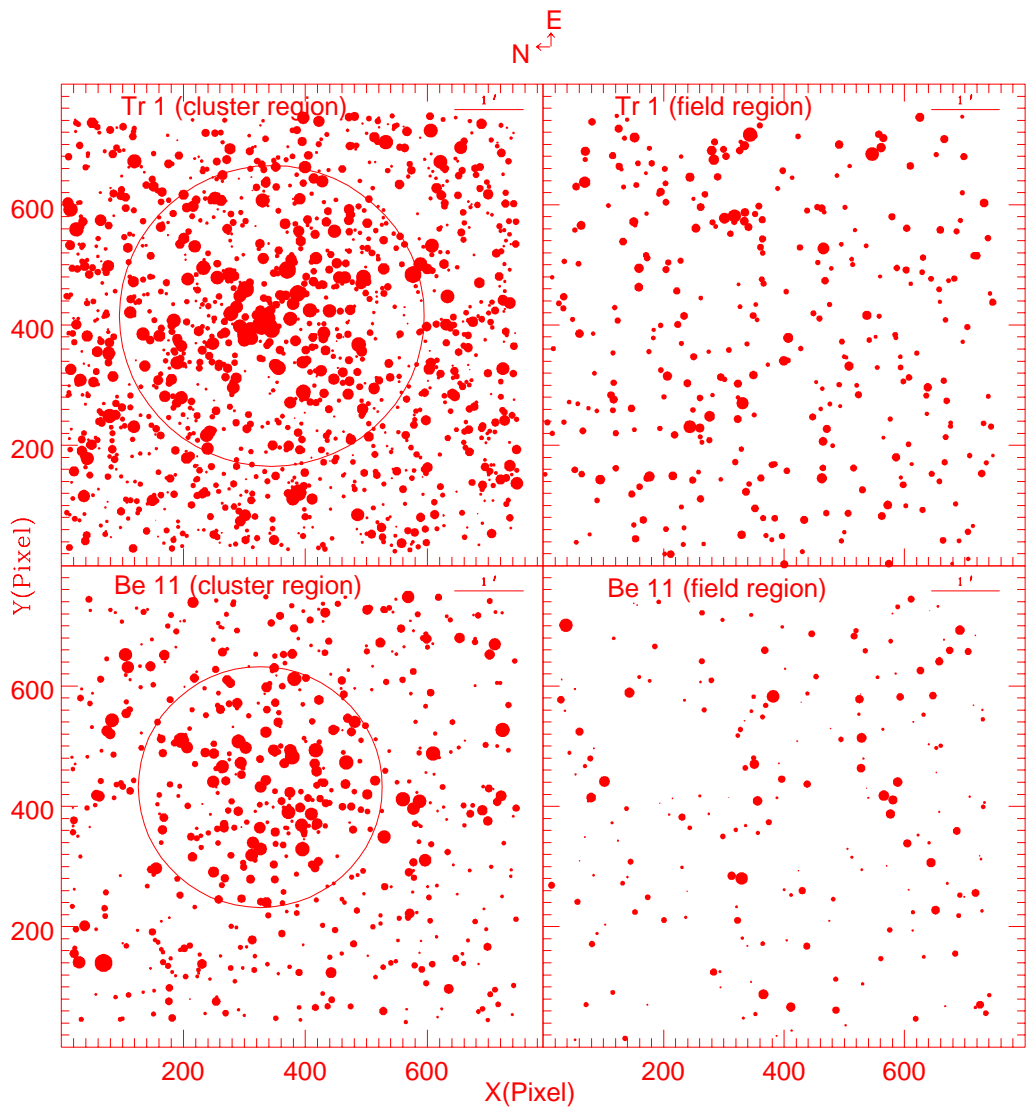,width=15cm,height=15cm}
\caption{Identification maps for the cluster and field regions of Tr 1
and Be 11. The (X, Y) coordinates are in pixel units corresponding
to 0$^{\prime\prime}$.72 on the sky. East is up and north is left. Filled
circles of different sizes represent brightness of the stars. Smallest size
denotes stars of $V$$\sim$21 mag. Open circles in the clusters region represent
the cluster size.}
\end{figure*}

The $UBV$ Johnson and $RI$ Cousins observations were obtained, using 
2K$\times$2K
CCD system at the f/13 Cassegrain focus of the Sampurnanand 104-cm telescope of
the
State Observatory, Nainital during November 2000. Details of the 
observations are
given in Table 2. Each pixel of 2048$\times$2048 size CCD corresponds to
 a square of size 0$^{\prime\prime}$.36 on the sky. In order to
improve the S/N ratio, the observations were taken in binning mode of 2$\times$2
 pixel.
The entire chip covers a field of 12$^{\prime}$.3$\times$12$^{\prime}.3$ though
because of the smaller filter
sizes, we could image only $\sim$ 8$^{\prime}$.6$\times$8$^{\prime}$.6 region. 
The
read-out noise for the system is 5.3 e$^{-}$ with a gain of 10 e$^{-}$/ADU.
In each passband, only one short but 2 to 3 deep exposures were taken
 for the cluster region so that accurate photometric measurements can be
obtained
for faint stars. For removing field star contamination, we also observed
field regions in the $UBV$($RI$)$_{C}$ passband situated $\sim$ 15$^{\prime}$
 south for both the clusters.
 Fig. 1 show the identification
 maps of imaged cluster and field regions of Tr 1 and Be 11.
 For calibration, we
observed 11 Landolt (1992) standard stars covering a range in brightness
(11.0$<$$V$$<$15.0) as well as in colour (0.16$<$$(V-I)$$<$2.08). Flat field 
exposures ranging from 20 to 60 sec in each filter were made on the twilight 
sky. A number of biases were also taken during the observing runs.\\

The CCD data frames were reduced using computing facilities available at 
the State Observatory,
 Nainital. Initial processing of the data frames were done in the
usual manner using the IRAF data reduction package. Different cleaned frames of
the same field in the same filter were co-added. Photometry of co-added frames 
was
carried out using DAOPHOT software (Stetson 1987). PSF was obtained for each
frame using several uncontaminated stars. In those cases where brighter stars
are saturated on deep exposure frames, their magnitudes have been taken only
from the short exposure frames. Wherever more than one measurement is available
 in a passband for a star, the final magnitude is an average of the individual
measurements and its error is the ALLSTAR error of the average. When only one
measurement is available, the error is taken to be the output of ALLSTAR.\\

\begin{table*}
\centering
\begin{minipage}{150mm}
\caption{Log of CCD observations. N denotes the number of stars
measured in different passbands.}
\begin{center}
\begin{tabular}{ccccc}
\hline
Region & Filter  &Exposure Time &Date&N\\
&&(in seconds)   & &\\
\hline
Trumpler 1 - cluster&$U$&1800$\times$2, 300$\times$1&19/20 Nov 2000&524\\
$\alpha_{2000}=01^{h}35^{m}40^{s}$&$B$&1200$\times$3, 240$\times$1&,,&1283\\
$\delta_{2000}=+61^{d}17^{\prime}20^{\prime\prime}$&$V$&900$\times$3, 120$\times$1
&,,&1451\\
&$R$&480$\times$3, 60$\times$1&,,&1630\\
&$I$&240$\times$3, 60$\times$1&,,&1710\\
Trumpler 1 - field&$U$&1200$\times$1&20/21 Nov 2000&350\\
$\alpha_{2000}=01^{h}35^{m}40^{s}$&$B$&900$\times$1&,,&915\\
$\delta_{2000}=+61^{d}02^{\prime}20^{\prime\prime}$&$V$&600$\times$1&,,&1190\\
&$R$&300$\times$1&,,&1346\\
&$I$&300$\times$1&,,&1585\\
\hline
Berkeley 11 - cluster&$U$&1800$\times$2, 300$\times$1&18/19 Nov 2000&295\\
$\alpha_{2000}=04^{h}20^{m}36^{s}$&$B$&1200$\times$3, 240$\times$1&,,&650\\
$\delta_{2000}=+44^{d}55^{\prime}58^{\prime\prime}$&$V$&900$\times$3, 180$\times$1
&,,&750\\
&$R$&600$\times$3, 120$\times$1&,,&930\\
&$I$&300$\times$3, 60$\times$1&,,&1090\\
Berkeley 11 - field&$U$&900$\times$1&20/21 Nov 2000&90\\
$\alpha_{2000}=04^{h}20^{m}34^{s}$&$B$&600$\times$1&,,&299\\
$\delta_{2000}=+44^{d}40^{\prime}08^{\prime\prime}$&$V$&600$\times$1&,,&474\\
&$R$&300$\times$1&,,&676\\
&$I$&300$\times$1&,,&723\\
\hline
\end{tabular}
\end{center}
\end{minipage}
\end{table*}

The photometric calibration equations are determined by fitting least
square linear regression to the standard $UBVRI$ photometric indices as function
of observed instrumental magnitudes normalized for 1 second exposure time. The
following colour equations are obtained for the system.

\begin{center}
$\Delta$$(U-B)$ = (0.973$\pm$0.020)$\Delta$($u-b$)$_{0}$

$\Delta$$(B-V)$ = (1.117$\pm$0.014)$\Delta$($b-v$)$_{0}$

$\Delta$$(V-R)$ = (0.886$\pm$0.011)$\Delta$($v-r$)$_{0}$

$\Delta$$(V-I)$~ = (0.992$\pm$0.007)$\Delta$($v-i$)$_{0}$

$\Delta$$V$ = $\Delta$$v_{0}$ $-$ (0.060$\pm$0.024)$(V-I)$\\
\end{center}
      where $\Delta$ denotes differential values; $(U-B)$, $(B-V)$, $(V-R)$, $(V-I)$ and $V$ are standard values
taken from Landolt (1992) and $(u-b)$$_{0}$, $(b-v)$$_{0}$, $(v-r)$$_{0}$ $(v-i)
$$_{0}$ and $v_{0}$ are the instrumental CCD aperture magnitudes and colours
corrected for atmospheric extinction. The atmospheric extinction coefficients
are 0.59$\pm$0.07, 0.37$\pm$0.02, 0.29$\pm$0.06, 0.16$\pm$0.03 and 0.13$\pm0.02$
 for $U$,
 $B$, $V$, $R$ and $I$ respectively. The errors in the colour coefficients  are obtained from the deviation of data
points from the linear relation. These equations are used to standardize the
CCD instrumental magnitudes of
both cluster and field regions. To establish the local standards, we selected
several isolated stars in the observed regions and used the DAOGROW programme
for the construction of an aperture growth curve required for determining the
difference between aperture and profile - fitting magnitudes. These differences
and differences in exposure times and atmospheric extinctions are used in 
evaluating zero points for the reference frames. The zero points are uncertain 
by $\sim$ 0.02 mag in $V$; $\sim$ 0.01 mag in $(B-V)$, $(V-R)$, $(V-I)$ and
$\sim$ 0.03 mag in $(U-B)$. Other factors contributing to the photometric
uncertainty are described recently by Moitinho (2001). Amongst them, the
internal errors estimated on the S/N
ratio of the stars as output of the ALLSTAR mainly produce the scatter in the
various CC and CM diagrams of the clusters. They are given in Table 3 as a
function of brightness for the cluster region. The errors become large ($\ge$0.1
 mag) for stars fainter than $V$=20 mag, so the measurements should be 
considered unreliable below this magnitude.\\

\begin{figure}
\psfig{file=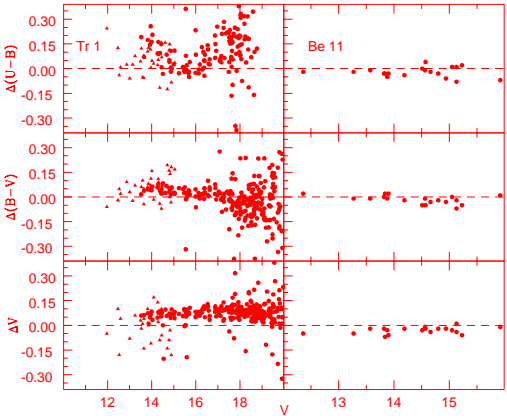,width=8cm,height=8cm}
\caption{A comparison of present photometry with CCD and photoelectric
data of Tr 1 and Be 11. For Tr 1, filled circles represent the CCD data of Phelps
 \& Janes (1994) and triangles represent the photoelectric data of Joshi \& Sagar
 (1977). For Be 11 filled circles represent the photoelectric data of Jackson et
al. (1980).}
\end{figure}

     The (X, Y) pixel coordinates as well as the $V$, $(U-B)$, $(B-V)$, $(V-R)$
and $(V-I)$ magnitudes of a sample of stars observed in the cluster and field 
regions of Tr 1 and Be 11 are listed in Tables 4 and 5 respectively. Only formats
of the tables are presented here; the full tables are available in the 
electronic version of the article on Synergy, on the open star cluster data 
base web
site at {\it http://obswww.unige.ch/webda/}, and also with the authors. In order
 to avoid introducing a new numbering system in Tr 1, we adopt the numbers
from the data base given by Phelps \& Janes (1994). Stars not observed earlier 
have a number starting with 1001 in Tr 1.
\begin{table}
 \centering
\caption{Internal photometric errors as a function of brightness. $\sigma$ is 
the standard deviation per observation in magnitude.}

\begin{tabular}{cccccc}
\hline
Magnitude range& $\sigma$$_{U}$&$\sigma$$_{B}$&$\sigma$$_{V}$  &$\sigma$$_{R}$ &
$\sigma$$_{I}$\\
\hline
$\le$12.0&0.012&0.001&0.005&0.003&0.011\\
12.0 - 13.0&0.013&0.004&0.008&0.008&0.013\\
13.0 - 14.0&0.019&0.006&0.010&0.009&0.019\\
14.0 - 15.0&0.021&0.007&0.014&0.012&0.014\\
15.0 - 16.0&0.023&0.012&0.015&0.013&0.014\\
16.0 - 17.0&0.025&0.012&0.017&0.015&0.020\\
17.0 - 18.0&0.043&0.024&0.026&0.025&0.030\\
18.0 - 19.0&0.083&0.054&0.055&0.051&0.060\\
19.0 - 20.0&0.086&0.095&0.102&0.091&0.113\\
20.0 - 21.0&0.158&0.281&0.356&0.346&0.264\\
\hline
\end{tabular}
\end{table}
\subsection{Comparison with the previous photometries}
We compare the present data with the published CCD and photoelectric data.
Table 6 represents
the average differences in the sense present minus others along with their
standard deviations. The differences $\Delta$ in mag and colours are plotted in
Fig 2. Fig 2 and Table 6 indicates that \\

present CCD data for Tr 1 show a constant zero point 
offset of $\sim$ 0.08 mag in $\Delta$V with that of given by Phelps \& Janes (1994). 
A weak linear dependence of $\Delta (B-V)$ is 
observed on brightness. The $\Delta (U-B)$ values, on the other hand, show decreasing trend upto 
$V \sim $ 16.0 mag but increase for fainter stars. As $UBV$ photoelectric data 
is in good agreement with the present CCD data for both the clusters, we suspect 
calibration problem with Phelps \& Janes (1994) CCD data.\\
\begin{table*}
 \centering
 \begin{minipage}{140mm}
\caption{CCD relative (X, Y) positions and $V$, $(U-B)$, $(B-V)$, $(V-R)$ and $(V-I)$ photometric magnitudes of few stars,
as a sample measured in the cluster and field region of the cluster Tr 1. 
In the cluster region, stars observed earlier have numbering system of Phelps 
\& Janes (1994) taken from the cluster data base, while numbering of stars 
observed for the first time by us start with 1001 in the column 1. The last 
column represent the photometric membership informations where m and nm
represent the member and non-member stars. In the field region, stars are 
numbered in the increasing order of X value.}

\begin{tabular}{ccccccccc}
\hline
Star&X&Y&$V$&$(U-B)$&$(B-V)$&$(V-R)$&$(V-I)$&Mem\\
&(pixel)&(pixel)&(mag)&(mag)&(mag)&(mag)&(mag)\\
\hline
&&&&Tr 1 cluster region&&&\\
1001&    97.64&    229.68&  19.80&      *&  1.21&   0.72&   1.49&  nm\\
1002&   356.27&    238.04&  18.67&   0.42&  1.16&   0.66&   1.35&   m\\
1003&   338.42&    237.33&  19.54&      *&  1.31&   0.78&   1.54&   m\\
1004&   309.76&    235.40&  18.04&   0.48&  1.17&   0.66&   1.31&   m\\
1005&   155.45&    227.56&  17.66&   0.61&  1.08&   0.65&   1.26&  nm\\
&&&&Tr 1 field region&&&&\\
1&      3.32&    151.93&   19.36&      *&  1.20&  0.68&  1.46&\\
2&      6.93&    238.09&   18.45&      *&  1.23&  0.67&  1.45&\\
3&     14.91&    479.27&   19.46&      *&  1.21&  0.73&  1.60&\\
4&     15.22&    219.91&   18.94&      *&  1.37&  0.67&  1.52&\\
5&     16.93&    360.70&   19.06&      *&  1.30&  0.65&  1.46&\\
\hline
\end{tabular}
\end{minipage}
\end{table*}

\begin{figure}
\hspace{2.0cm}\psfig{file=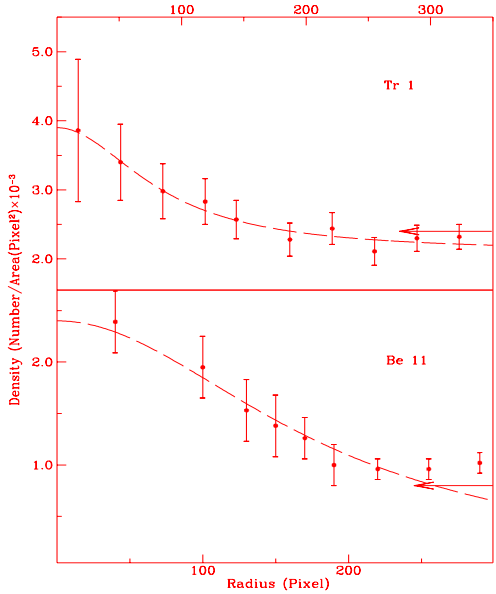,width=9.5cm,height=8.5cm}
\caption{Radial density profile for Tr 1 and Be 11. The length
of the errorbar denotes errors resulting from sampling statistics 
(=$\frac{1}{\sqrt{N}}$ where N is the number of stars used in the density 
estimation at that point). Dotted curves represent fitted profile and arrows 
represent the level of field star densities.}
\end{figure}

\begin{table*}
 \centering
 \begin{minipage}{140mm}
\caption{CCD relative (X, Y) positions and $V$, $(U-B)$, $(B-V)$, $(V-R)$ and
$(V-I)$ photometric magnitudes of few stars, as a sample measured in the 
cluster and field regions of the cluster Be 11. Stars are numbered in 
increasing order of X value. The last column indicates the 
photometric member and non-member stars of the cluster.}

\begin{tabular}{ccccccccc}
\hline
Star&X&Y&$V$&$(U-B)$&$(B-V)$&$(V-R)$&$(V-I)$&Mem\\
&(pixel)&(pixel)&(mag)&(mag)&(mag)&(mag)&(mag)\\
\hline
&&&Be 11 cluster region&&&&\\
142& 311.95&    318.82&   14.78&   0.61&   0.99&   0.48&   1.04&    m\\
143& 312.78&    178.07&   16.97&   0.77&   1.30&   0.63&   1.33&    m\\
144& 313.98&    339.75&   15.11&   0.43&   0.91&   0.45&   0.97&    m\\
145& 316.18&    128.17&   18.53&      *&   1.54&   0.77&   1.65&    nm\\
146& 317.11&    151.06&   19.43&      *&   1.81&   0.92&   1.88&    m\\
&&&Be 11 field region&&&&\\
1&   14.00&    268.90&   18.02&      *&   2.10&   1.15&   2.40&\\
2&   29.36&    577.23&   17.64&   0.60&   1.10&   0.62&   1.29&\\
3&   30.76&    610.82&   20.98&      *&      *&   0.84&   1.65&\\
4&   36.19&    558.98&   20.62&      *&      *&   0.77&   1.56&\\
5&   37.84&    701.05&   14.40&   0.63&   1.10&   0.50&   1.06&\\ 
\hline
\end{tabular}
\end{minipage}
\end{table*}

\begin{table*}
% \centering
 \begin{minipage}{140mm}
 \caption{Comparison of our photometry with others for the cluster 
Tr 1 and Be 11. The difference ($\Delta$) is always in the sense present minus 
comparison data. The mean and standard deviations in magnitude are based on 
N stars. Few deviated points are not included in the average determination.}
\begin{tabular}{cccccc}
\hline
Cluster&Comparison data&$V$ range&$<\Delta$$V>$&$<\Delta(B-V)$$>$&$<\Delta(U-B)>$\\
&&&Mean$\pm$$\sigma$(N)&Mean$\pm$$\sigma$(N)&Mean$\pm$$\sigma$(N)\\
\hline
Tr 1&Phelps \& Janes (1994)&$<$ 14.0&\hspace{-0.15cm}0.05$\pm$0.02(7)&\hspace{-0.15cm}0.03$\pm$0.01(6)&\hspace{-0.15cm}0.16$\pm$0.06(6)\\
&&14.0 $-$ 15.0&0.06$\pm$0.02(17)&0.04$\pm$0.02(17)&0.07$\pm$0.07(14)\\
&&15.0 $-$ 16.0&0.05$\pm$0.06(19)&0.02$\pm$0.02(16)&0.02$\pm$0.07(15)\\
&&16.0 $-$ 16.5&0.08$\pm$0.02(12)&0.01$\pm$0.03(12)&0.03$\pm$0.05(12)\\
&&16.5 $-$ 17.0&0.08$\pm$0.03(10)&0.03$\pm$0.03(10)&0.11$\pm$0.07(10)\\
&&17.0 $-$ 17.5&0.08$\pm$0.03(14)&0.01$\pm$0.08(13)&0.13$\pm$0.10(11)\\
&&17.5 $-$ 18.0&0.08$\pm$0.04(37)&\hspace{-0.2cm}$-0.03\pm0.05$(37)&0.12$\pm$0.09(29)\\
&&18.0 $-$ 18.5&0.08$\pm$0.05(30)&\hspace{-0.2cm}$-0.03\pm0.09$(30)&0.10$\pm$0.11(13)\\
&&18.5 $-$ 19.0&0.07$\pm$0.03(43)&\hspace{-0.2cm}$-0.02\pm0.07$(43)&\\
&Joshi \& Sagar (1977)&12.0 $-$ 13.0&\hspace{-0.2 cm}$-0.01\pm0.14$(3)&\hspace{-0.2cm}0.03$\pm$0.04(4)&\hspace{-0.2cm}0.03$\pm$0.07(4)\\
&&13.0 $-$ 14.0&0.00$\pm$0.06(10)&0.05$\pm$0.04(12)&0.04$\pm$0.07(12)\\
&&14.0 $-$ 15.0&\hspace{-0.10cm}0.03$\pm$0.09(7)&0.07$\pm$0.08(17)&0.03$\pm$0.08(17)\\

Be 11&Jackson et al. (1980)&13.0 $-$ 14.0&\hspace{-0.20cm}$-0.05\pm0.02$(6)&\hspace{-0.20cm}0.00$\pm$0.01(6)&\hspace{-0.20cm}$-0.03\pm0.01$(6)\\
&&14.0 $-$ 15.0&\hspace{-0.20cm}$-0.03\pm0.01$(7)&\hspace{-0.20cm}$-0.03\pm0.01$(7)&\hspace{-0.20cm}$-0.02\pm0.03$(7)\\
&&15.0 $-$ 16.0&\hspace{-0.20cm}$-0.03\pm0.02$(5)&\hspace{-0.2cm}$-0.02\pm0.03$(5)&\hspace{-0.2cm}$-0.02\pm0.04$(5)\\
\hline
\end{tabular}
\end{minipage}
\end{table*}
\subsection {Near - IR data}
The near-IR $JHK$ data are taken from the digital Two Micron
All Sky Survey (2MASS) available at web site {\it http://www.ipac.caltech.edu/2MASS/}. 2MASS is uniformly scanning the entire sky in three near-IR bands 
$J$(1.25 $\mu$m), $H$(1.65 $\mu$m) and $K$$_{s}$(2.17 $\mu$m) with two highly - automated 1.3-m
telescopes equipped with a three channel camera, each channel consisting of a
256$\times$256 array of HgcdTe detectors. The photometric uncertainty of the 
data is $<$ 0.155 mag with $K$$_s$ $\sim$ 16.5 mag photometric completeness. 
The $K$$_s$ magnitudes are converted into $K$ magnitude following Persson 
et al. (1998). Further details about the 2MASS and 2MASS data are available at
{\it http://www.ipac.caltech.edu/2mass/releases/second/doc/
explsup.html}. The 
$JHK$$_s$ data are available for 200 stars in the observed area of Be 11 
cluster. Among them only 179 stars are common with our optical data.

\section {Data analysis}

\subsection {Cluster radius and radial stellar surface density}

   The first step to determine the cluster radius is to find its center in the
image. The center of the cluster is determined iteratively by calculating
average X and Y positions of the stars within 300 pixels from an eye estimated
center, until they converged to a constant value. An error of a few tens of
pixels is
expected in locating the cluster center. The (X, Y) pixel coordinates
of the cluster centers obtained in this way are (345, 415) and (326, 432) for
Tr 1 and Be 11 respectively. The corresponding equatorial coordinates are given
 in Table 2. For determining the radial surface density of stars $\rho(r)$ in a
cluster, the
imaged area has been divided into a number of concentric circles with respect to
the above estimated center, in such a way that each zone contains a 
statistically significant number of stars. The number density of stars, 
$\rho_{i}$, in the i$^{th}$ zone has been calculated as $\rho_{i}$ = 
$\frac {N_{i}}{A_{i}}$,
where $N_{i}$ is the number of stars and $A_{i}$ is the area of the
i$^{th}$ zone. The density versus radius plots for Tr 1 and Be 11 are shown in
Fig 3. A clear radius-density gradient present in Fig 3 confirms the
existence of clustering. Following Kaluzny (1992), we describe the $\rho$(r) of
an open cluster as:
\begin{displaymath}
~~~~~~~~~~~~~~~~~~~ \rho(r) \propto \frac {f_0}{1+(r/r_{c})^2},
\end{displaymath}
where the cluster core radius r$_{c}$ is the radial distance at which the
value of $\rho(r)$ becomes half of the central density ${f_0}$. We fit this
function to the observed data points of each cluster and use $\chi^{2}$ 
minimization technique to determine $r_{c}$ and other constants. As can be 
seen in Fig 3, the fitting of the function is satisfactory. The values of core radii derived in
this way are 85$\pm$10 and 180$\pm$10 pixels for Tr 1 and Be 11 respectively. Fig
3 also separate the
cluster region
from the surrounding field region. The field star density thus obtained are 2.2$
\times$10$^{-3}$ and 1.0$\times$10$^{-3}$ stars/pixel$^{2}$ for Tr 1 and Be 11 
respectively. The
radial distribution of stars in Tr 1 and Be 11 indicates that the extent of the 
cluster is about
250 and 200 pixels respectively which correspond to $\sim$ 3.0 and 2.4 in 
arcmin. The cluster sizes are thus a few times larger than the corresponding
core sizes which is in agreement with the findings of Nilakshi et al. (2002).

\subsection {Apparent colour-magnitude diagrams of cluster and field regions}

      The apparent CM diagrams generated from the present data for the
 Tr 1 and Be 11 clusters and their field regions are displayed in Fig 4. The CM
diagrams
 extend up to $V$ $\sim$ 21 mag except in $V$, $(U-B)$ diagram where it is only
 up to $V$ $\sim$ 18 mag. A well defined cluster MS contaminated by field stars
 is clearly visible in all CM diagrams. The field star contamination increases
with decreasing brightness. The cluster sequence fainter than $V$ $\sim$ 17 mag
have larger scatter. This may be due to photometric errors as
well as field star contamination. It is difficult to separate field stars from
the cluster members only on the basis of their closeness to the main populated
area of the CM diagrams, because field stars at cluster distance and reddening
also occupy this area (see Fig 4). For the separation of cluster members from
the field stars, precise proper motion and/or radial velocity measurements of
these stars are required. In the absence of such data, we use photometric
criterion for separating obvious field stars. A star is considered as a 
non-member if it lies outside the cluster sequence in at least one CC or 
CM diagrams. From the $V$, $(V-I)$ diagram of the field 
region, statistically expected number of field stars among the
photometric cluster members has been given in Table 7. The frequency
distribution of the field star contamination in different part of the CM
diagram can be estimated from the Table 7. It is thus clear that all
photometric probable members can not be cluster members and non-members should
be subtracted in the studies of cluster MF etc. However, probable members
located within a cluster radius from its center can be used to determine the
cluster parameters, as they have relatively less field star contamination and 
this has been done in the sections to follow.

\begin{figure*}
\psfig{file=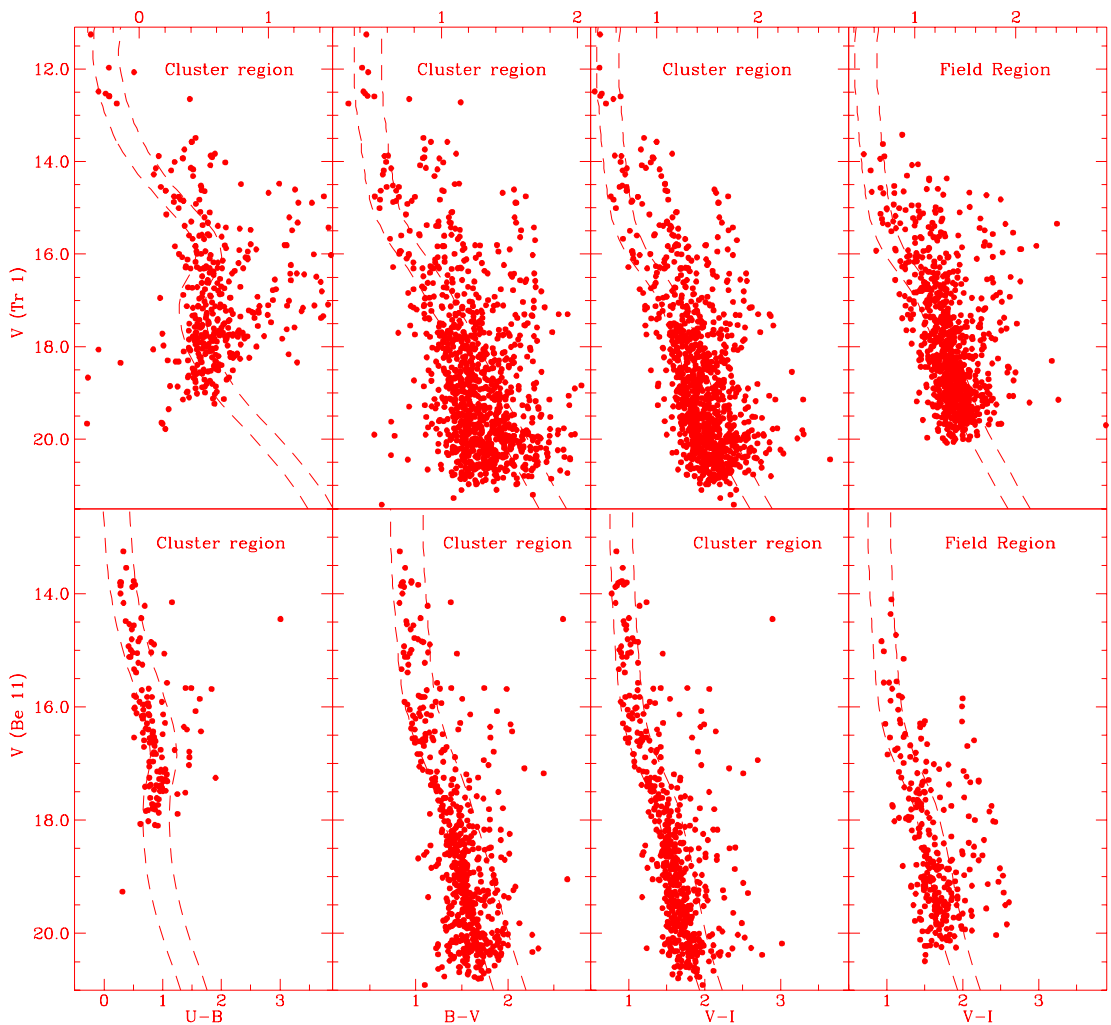,width=15cm,height=15cm}
\caption{The $V$, $(U-B)$; $V$, $(B-V)$ and $V$, $(V-I)$ 
diagrams for the stars observed by us in Tr 1 and Be 11 cluster regions and the 
$V$, $(V-I)$ CM diagram of the corresponding field regions. Dotted lines represent 
the blue and red envelope of the cluster MS.}
\end{figure*}

\subsection {Interstellar extinction in the direction of clusters}
To estimate interstellar extinction in the direction of the clusters, we
plot in Fig 5 $(U-B)$ versus $(B-V)$ diagrams of the sample stars.
Adopting the slope of reddening line
$E(U-B)/E(B-V)$ as 0.72, we fit the intrinsic zero-age main-sequence (ZAMS)
given by Schmidt-Kaler (1982) to the MS stars of spectral type
earlier than A0. This gives a mean value of $E(B-V)$ = 0.60$\pm$0.05 mag for the
 cluster
Tr 1 and 0.95$\pm$0.05 mag for the cluster Be 11.
 Our reddening estimates for the imaged
 region agree fairly well with the values estimated earlier by others (see Table
 1).\\

\begin{figure}
\psfig{file=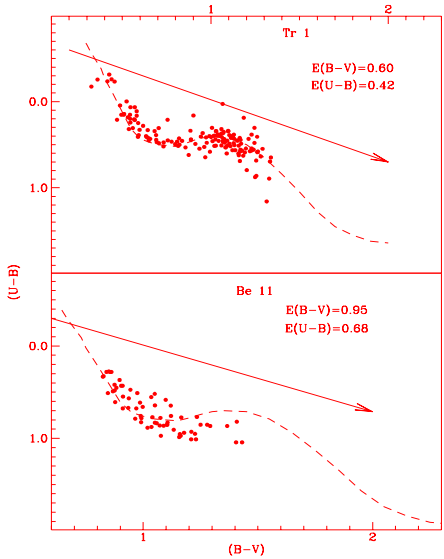,width=8.5cm,height=8cm}
\caption{The $(U-B)$ versus $(B-V)$ diagrams of the stars in cluster region observed
by us for Tr 1 and Be 11. The continuous straight line represents slope 0.72
and direction of the reddening vector. The dotted curve represents the
locus of Schmidt - Kaler's (1982) ZAMS fitted for the marked values of colour
excesses.}
\end{figure}

      For determining the nature of interstellar extinction law in the direction
of clusters, we used the stars having spectral type earlier than A0. This has
been selected from their location in the $(U-B)$ versus $(B-V)$ and apparent 
CM diagrams which reveals that bright stars ($V$$<$ 15.0 mag) with
$(B-V)$$<$0.90 mag in Tr 1 and with $(B-V)<1.30$ mag in Be 11 are wanted 
objects. For these stars, their intrinsic colours have been determined using 
either the spectral
type (available only for 15 stars in Be 11) taken from the open cluster data
base (cf. Mermilliod 1995) or $UBV$ photometric Q-method (cf. Johnson \& Morgan
1953; Sagar \& Joshi 1979) and the calibrations given by Caldwell et al. (1993)
for $(U-B)_{0}$, $(V-R)_{0}$ and $(V-I)_{0}$ with $(B-V)_{0}$. The mean values
of the colour excess ratios derived
in this way are
listed in Table 8 for both the clusters. They indicate that the law of interstellar 
extinction in the direction of the clusters under discussion is normal.
\begin{table*}
 \centering
  \begin{minipage}{140mm}
   \caption{Frequency distribution of the stars in the $V$, $(V-I)$
 diagram of the cluster and field regions. $N_{B}$, $N_{S}$ and $N_{R}$ denote
the number of stars in a magnitude bin blueward, along and redward of the
cluster sequence respectively. $N_{C}$ (difference between the $N_{S}$ value
of cluster and field regions) denotes the statistically expected number of
cluster members in the magnitude bin.}
\vspace{0.7cm}
\begin{tabular}{|c|ccc|ccc|c|ccc|ccc|c|}
\hline
&\multicolumn{7}{|c|}{Tr 1} & \multicolumn{7}{c|}{Be 11} \\
\cline{2-15}
V range &\multicolumn{3}{|c|}{Cluster region} & \multicolumn{3}{|c|}{Field 
region}& &
\multicolumn{3}{|c|}{Cluster region} & \multicolumn{3}{|c|}{Field region}&  \\
&$N_{B}$&$N_{S}$&$N_{R}$&$N_{B}$&$N_{S}$&$N_{R}$&$N_{C}$&$N_{B}$&$N_{S}$&$N_{R}$
&$N_{B}$&$N_{S}$ &$N_{R}$ &$N_{C}$ \\
\hline
13 - 14 & 0 & 8& 8 & 1 & 3 & 1 & 5 & 0  & 10  &0  & 0 & 0 & 0  & 10 \\
14 - 15 & 0 & 20&25 & 0 & 3 & 21 & 17 & 0  & 16 &4  & 0 & 4 & 0  & 12  \\
15 - 16 & 0 & 22&40 & 1 & 5& 60& 17 & 0  & 26 &10 & 0 & 8 & 1  & 18 \\
16 - 17 & 3 & 37&53 &4 & 25& 80& 12 & 0  & 48 &14 & 6 & 9 & 10 & 39 \\
17 - 18 & 18&97& 75&24 & 74& 74& 23& 0  & 68 &15 &22 &16 & 14 & 52 \\
18 - 19 & 52&131&71&80 &110 & 52& 21& 14 & 142 &19 &33 & 22 & 14 & 120\\
19 - 20 &130&139&51&105&110&28&29& 45 & 159 &14 &68 &34 & 14 & 125 \\
\hline
\end{tabular}
\end{minipage}
\end{table*}
\subsubsection {Presence of non-uniformity in $E(B-V)$}
In order to see the extent of non-uniform extinction in the clusters under
 study, we plot histograms of $E(B-V)$ in Fig 6. They indicate that there
is a range of $\sim$ 0.4 mag in $E(B-V)$ values of both the cluster stars with
peak around $E(B-V)$ = 0.65 and 0.95 mag for Tr 1 and Be 11 respectively.
\begin{figure}
\hspace{1.0cm}\psfig{file=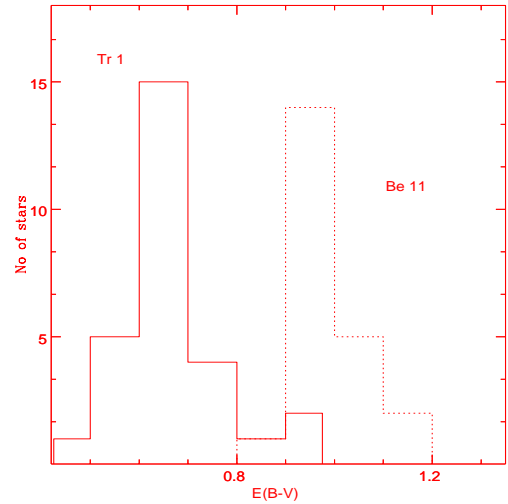,width=9cm,height=8cm}
\caption{Solid and dotted lines represent the histograms of $E(B-V)$ for
Tr 1 and Be 11
respectively. They indicate presence of non - uniform extinction in both
clusters.}
\end{figure}
To determine the presence of non-uniform extinction in a cluster, we
calculate the value of $\Delta$$E(B-V)$ = $E(B-V)$$_{max}$$-$$E(B-V)$$_{min}$, 
where $E(B-V)$$_{max}$ and $E(B-V)$$_{min}$ are determined on the basis of 
respectively,
the five highest and five lowest $E(B-V)$ values of the MS cluster members.
In this way we can find the values of $\Delta$$E(B-V)$ and these values are
listed in Table 9. As the factors other than non-uniform extinction like stellar
evolution, stellar duplicity, stellar rotation, difference in chemical
composition, dispersion in ages, distances, and
uncertainty in photometric data (cf. Burki 1975, Sagar 1987, Yadav \& Sagar 2001) can produce
a maximum dispersion in $E(B-V)$ of $\sim$ 0.11 mag for MS members, we
consider the presence of non-uniform extinction in the both cluster regions, as
the observed values of $\Delta$$E(B-V)$ for their MS stars are much more than
0.11 mag. In order to see whether it is due to presence of varying amount of
matter inside them or due to any other regions, we studied below the variation
of $E(B-V)$ with spatial position of stars in the clusters.

\subsubsection {Spatial variation of $E(B-V)$}
 To study the spatial variation of reddening in terms of $E(B-V)$ across the 
cluster region, we divide the cluster field into equal area of small boxes of size
1$^{\prime}$.0$\times$1$^{\prime}$.0. The
positional variation of $E(B-V)$ is shown in Table 10 for both the clusters.
An inspection of Table 10 indicates that $E(B-V)$ does not show any systematic 
variation with position in both the clusters. However, it varies randomly which may be due to the 
random distribution of gas and dust within the clusters.

\begin{table}
 \centering
 \caption{A comparison of the colour excess ratios with $E(B-V)$ for star
clusters with the corresponding values for the normal interstellar extinction
law given by Cardelli et al. (1989).}

\begin{tabular}{cccc}
\hline
Objects&$\frac{E(U-B)}{E(B-V)}$&$\frac{E(V-R)}{E(B-V)}$&$\frac{E(V-I)}{E(B-V)}$\\
\hline
Normal interstellar&0.72&0.60&1.25\\
Tr 1&0.72$\pm$0.03&0.54$\pm$0.02&1.13$\pm$0.05\\
Be 11&0.71$\pm$0.02&0.45$\pm$0.02&1.04$\pm$0.04\\
\hline
\end{tabular}
\end{table}

\begin{table}
\caption {The values of $E(B-V)$$_{min}$, $(B-V)$$_{max}$ and 
$\Delta$$E(B-V)$ (see text).}

\begin{tabular}{cccc}
\hline
Cluster&$E(B-V)$$_{min}$&$E(B-V)$$_{max}$&$\Delta$$E(B-V)$\\
&(mag)&(mag)&(mag)\\
\hline
Tr 1&0.55&0.85&0.30\\
Be 11&0.91&1.06&0.15\\
\hline
\end{tabular}
\end{table}

\begin{figure}
\psfig{file=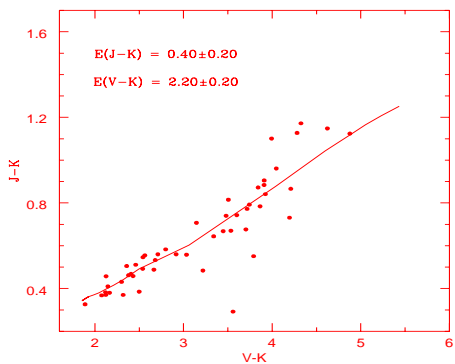,width=8.5cm,height=7cm}
\caption{The $(J-K)$ versus $(V-K)$ colour - colour diagram of all the stars
 which are common in $V$ and $JHK$ data within the cluster radius for the 
cluster Be 11. The solid line is a ZAMS fitted for the marked values of colour 
excesses.}
\end{figure}

\subsubsection {Interstellar extinction in near - IR}

In order to derive the interstellar extinction for Be 11 in near - IR region, we
have combined optical and infrared data. There are 47 common stars within the 
cluster radius. We construct in Fig. 7 ($J-K$) vs ($V-K$) diagram for the stars 
within the cluster radius and fit a ZAMS for metallicity Z = 0.02 taken from 
Schaller et al. (1992). This yields $E(J-K)$ = 0.40$\pm$0.20 mag and $E(V-K)$ = 
2.20$\pm$0.20 mag, which corresponds to a ratio $\frac{E(J-K)}{E(V-K)}$ = 0.18$\pm$0.30 
in good agreement with the normal interstellar extinction value 
0.19 suggested by Cardelli et al. (1989). However, scattering is larger due to the 
error size in $JHK$ data.

\begin{table*}
\begin{minipage}{140mm}
\caption{Spatial variation of $E(B-V)$ across the cluster Tr 1 and Be 11.
 The mean values of $E(B-V)$ with their standard deviation in mag in 1$^{\prime}
$$\times$1$^{\prime}$ areas are indicated in the appropriate boxes, with the 
number of stars used for this purpose given in brackets. The $\Delta$$\alpha$ and
$\Delta$$\delta$ values are in arcmin relative to the cluster center given in
Table 2.}
\begin{center}
\begin{tabular}{|c|ccccc|}
\hline
$\Delta$$\alpha$$\rightarrow$~~ &&&Tr 1&&\\
$\Delta\delta$ $\downarrow$&$-3$ to $-2$& $-2$ to $-1$&$-1$ to 0&0 to 1&1 to 2\\
\hline
$-3$ to $-2$&-&0.60&-&-&-\\
&&(1)&&&\\
$-2$ to $-1$&-&0.60&-&0.73$\pm$0.21&-\\
&&(1)&&(3)&\\
$-1$ to 0&0.66&0.61&0.60$\pm$0.09&0.58&0.53\\
&(1)&(1)&(5)&(1)&(1)\\
0 to 1&-&0.67$\pm$0.07&0.65$\pm$0.04&0.85$\pm$0.20&0.67$\pm$0.07\\
&&(2)&(3)&(2)&(3)\\
1 to 2&-&0.68&-&-&0.64\\
&&(1)&-&-&(1)\\
2 to 3&-&0.79&0.71&-&-\\
&&(1)&(1)&&\\
&&&&&\\
&&&Be 11&&\\
$-2$ to $-1$&-&-&1.01$\pm$0.02&1.00$\pm$0.01&-\\
&&&(2)&(2)&\\
$-1$ to $0$&-&-&0.95&0.95$\pm$0.05&1.03\\
&&&(1)&(2)&(1)\\
$0$ to 1&-&1.05$\pm$0.20&0.98$\pm$0.04&0.94$\pm$0.05&0.99$\pm$0.01\\
&&(2)&(5)&(3)&(2)\\
1 to 2&-&-&-&-&-\\
&&&&&\\
2 to 3&-&&&0.94&-\\
&&&&(1)&\\
\hline
\end{tabular}
\end{center}
\end{minipage}
\end{table*}

\begin{figure}
\psfig{file=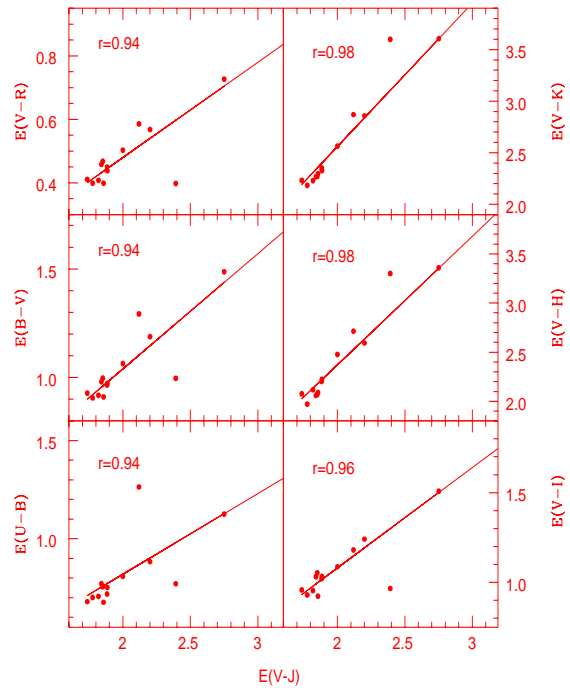,width=8.5cm,height=10cm}
\caption{The plot of $E(U-B)$, $E(B-V)$, $E(V-R)$ $E(V-I)$, $E(V-H)$ and
$(V-K)$ against $E(V-J)$ for Be 11. Solid line in each diagram
represents least square linear fit to the data points. The values of correlation
 coefficients are shown in the diagram.}
\end{figure}

\subsubsection {Extinction Law in Be 11}

    In order to see the nature of extinction law in Be 11, we plot the
colour
excess $E(U-B)$, $E(B-V)$, $E(V-R)$, $E(V-I)$, $E(V-H)$ and $E(V-K)$ against
$E(V-J)$ in Fig 8. For normalization we have used the colour excess $E(V-J)$ 
instead of $E(B-V)$,
 because $E(V-J)$ does not depend on properties like the chemical composition, shape, and structure, degree of alignment of the interstellar matter (Sagar
\& Qian 1990). Also, they are better measure of the total amount of interstellar
 extinction because of their larger values compared to either near-IR or optical
 colour excesses. A careful selection is needed in the case of colour excesses,
because the young clusters are embedded in emission nebulosity and also contains
 young stellar objects. In such environment, the blueing effect, ultra-violet
excess, circumstellar dust and gas shells etc, may be present in and around
the cluster stars. So, it is fruitful to use $V$ band rather than $U$ or $B$, 
because it is least affected in such cases. Similarly, in the near-IR, $J$ 
band is
 preferred, with a view to minimizing the contributions from the possible
presence of circumstellar material etc, around young stellar objects and also to
 choose a photometric band which most closely represents the emission from the
stellar photosphere. We have therefore estimated the colour excess ratios
relative to $E(V-J)$.

      In the Fig 8, solid line represents the least square linear fits to the
data points. The values of correlation coefficient (r) and fit indicate that 
the data points
are well represented by linear relation. The slopes of these straight lines 
represent reddening directions in the form of colour
excess ratios are given in Table 11. For comparison, 
the colour excess
ratios given by Cardelli et al. (1989) are also listed in the Table 11. The
present reddening directions agree well with those given for normal
interstellar extinction law.
   
\begin{figure}
\psfig{file=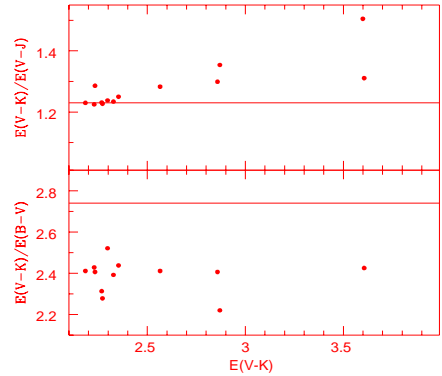,width=8.5cm,height=8cm}
\caption{The colour excess ratios $E(V-K)/E(V-J)$ (top panel) and $E(V-
K)/E(B-V)$ (bottom panel) as a function of $E(V-K)$ for Be 11. The horizontal
lines are drawn for the normal values of the colour excess ratios.}
\end{figure}

In the Fig 9, we have plotted ratios $E(V-K)$/$E(B-V)$ and $E(V-K)$/$E(V-J)$
 against $E(V-K)$. The horizontal line in the fig represents the value of the
ratio for the normal interstellar extinction law. In the case of normal 
extinction law, the ratio $E(V-K)$/$E(B-V)$ and $E(V-K)$/$E(V-J)$ remains the same
for all the values of $E(V-K)$ (Sagar \& Qian 1990). Least square linear fits to
 the data points gives
\begin{center}
$E(V-K)$/$E(B-V)$ = 0.10($\pm$0.03)$E(V-K)$ $+$ 2.13($\pm$0.31)~~~~r=0.29

$E(V-K)$/$E(V-J)$ = 0.03($\pm$0.02)$E(V-K)$ $+$ 1.18($\pm$0.05)~~~~r=0.44
\end{center}
         The values of r are $<$ 0.5. This indicates that the relations are not
statistically significant. This may therefore imply absence of anomalous 
interstellar
 extinction law toward the cluster Be 11.

      Whittet \& Breda (1980) suggested that, in the absence of complete data at
 long wavelengths, the approximation R = 1.1$E(V-K)$/$E(B-V)$ generally
used to deduce R is relatively insensitive to the reddening law adopted. We
have therefore used this relation to evaluate R. The average value of R = 2.60
$\pm$0.34 (sd), which is not too different from the value 3.1 for normal 
extinction law.
In the light of above analysis, we conclude that  interstellar extinction law is
 normal towards
Be 11 in agreement with our earlier result.

\subsection {Distance to the clusters}

    The distances of the clusters are derived by a ZAMS fitting procedure. We
have plotted intrinsic CM diagrams for Tr 1 and Be 11 in Fig 10. In order to
reduce field star contamination, we have used
only those probable cluster member stars which are within the cluster radius
from the cluster center. For plotting these diagrams, we have converted apparent
 $V$ magnitude and $(U-B)$, $(B-V)$, $(V-R)$ and $(V-I)$ colours into intrinsic
one using the available
individual values of $E(B-V)$
and following relations for $E(U-B)$ (cf. Kamp 1974; Sagar \& Joshi 1979), A$_{v
}$
and $E(V-I)$ (Walker 1987) and $E(V-R)$ (Alcala$^{\prime}$ et al. 1988).\\

$E(U-B)$ = [X + 0.05$E(B-V)$]$E(B-V)$

\vspace{0.3cm}
where X = 0.62 $-$ 0.3$(B-V)$$_{0}$ for $(B-V)$$_{0}$ $<$ $-$0.09

~~and~~~X = 0.66 + 0.08$(B-V)$$_{0}$ for$(B-V)$$_{0}$ $>$ $-0.09$\\

          A$_{v}$ = [3.06 + 0.25$(B-V)$$_{0}$ + 0.05$E(B-V)$]$E(B-V)$;\\
and      $E(V-R)$ = [E1 + E2E$(B-V)$]$E(B-V)$\\

where E1 = 0.6316 + 0.0713$(B-V)$$_{0}$\\
and E2 = 0.0362 + 0.0078$(B-V)$$_{0}$;\\

$E(V-I)$ = 1.25[1 + 0.06$(B-V)$$_{0}$ + 0.014$E(B-V)$]$E(B-V)$\\

For fainter stars, the average $E(B-V)$ value have been used for both the 
clusters since individual values are not known.\\

      In $V_{0}$, $(U-B)$$_{0}$ and $V_{0}$, $(B-V)$$_{0}$ diagrams, we 
fitted the ZAMS given by Schmidt-Kaler (1982) while the ZAMS given by Walker (1985)
was fitted in $V_{0}$, $(V-I)$$_{0}$ diagram. For $V_{0}$, $(V-R)$$_{0}$ 
diagram, we have calculated $(V-R)$$_{0}$ using its relation with $(B-V)_{0}$ 
given by Caldwell et 
al. (1993). The visual fit of the ZAMS to the bluest envelope of
the CM diagrams gives the mean values of
$(m-M)$$_{0}$ as 12.1$\pm$0.2 and 11.7$\pm$0.2 mag for Tr 1 and Be 11 
respectively. The distances to the
clusters should be considered reliable because they have been derived by fitting
the ZAMS over a wide range of the cluster MS. The distance modulus determined
above yields a distance of 2.6$\pm$0.10 Kpc to Tr 1 and of 2.2$\pm$0.10 Kpc to 
Be 11. For Tr 1, our values of distance
modulus is in good agreement with the value 12.10 given by Phelps \& Janes (1994
). However, it is slightly higher than the value 11.64 mag given by Joshi \&
Sagar (1977). The present determination of the distance modulus to the cluster
Be 11 agrees very well with the value of 11.7 mag given by Jackson et al. (1980)
.
\begin{table*}
 \begin{minipage}{140mm}
 \caption{A comparison of extinction law in the direction of Be 11 with normal 
extinction law given by Cardelli et al. (1989).}
\begin{center}
\begin{tabular}{cccccccc}
\hline
Source&$\frac{E(U-B)}{E(V-J)}$&$\frac{E(B-V)}{E(V-J)}$&$\frac{E(V-R)}{E(V-J)}$&$
\frac{E(V-I)}{E(V-J)}$&$\frac{E(V-H)}{E(V-J)}$&$\frac{E(V-K)}{E(V-J)}$&$\frac{E(
J-K)}{E(V-K)}$\\
\hline
Cardelli et al.&0.32&0.43&0.27&0.56&1.13&1.21&0.19\\
Colour excess ratio&0.42$\pm$0.13&0.53$\pm$0.03&0.29$\pm$0.03&0.56$\pm$0.04&1.31
$\pm$0.07&1.38$\pm$0.07&0.18$\pm$0.30\\
\hline
\end{tabular}
\end{center}
\end{minipage}
\end{table*}
\subsection {Gaps in MS}
The intrinsic CM diagrams of both the clusters Tr 1 and Be 11 
exhibit gaps at different points in the MS (see Fig 10). There seems to be a 
gap between 10.7 and 11.6 intrinsic $V$ mag in Tr 1. Another feature seen in 
this cluster 
MS is the deficiency of stars between 14.0 and 14.8 mag. In the 
case of Be 11, we noticed a gap between 12.4 and 12.9 intrinsic $V$ mag. The 
reality of 
the gaps in MS is tested by the method adopted by Hawarden (1971). The 
probability of finding such gaps to be accidental is 0.3\% for the gap 
between 10.7 and 11.6 in Tr 1 and 0.02\% for the gap between 12.4 and 12.9 in 
Be 11. The accidental probability is 12\% for the fainter gap in Tr 1. 
The very low values of probabilities indicate that the observed gaps are real. 
However, cause of such gaps are not well understood.
\subsection {Ages of the clusters}

      The ages of the clusters namely Tr 1 and Be 11 have been determined by
fitting the theoretical stellar evolutionary isochrones given by Schaller et al.
 (1992) in the corresponding CM diagrams (Fig 10). The isochrones are for Pop
I stars (X = 0.70, Y = 0.28, Z = 0.02) and include the effects of mass loss and
convective core overshooting in the model. The isochrone fitting to the main
sequence and brighter stars indicates that ages of the clusters Tr 1 and Be 11 
are 40$\pm$10 and 110$\pm$10 Myr respectively.
The present age estimate for Tr 1 and Be 11 are in good agreement with the
values given in Mermilliod (1995) (see Table 1). For Tr 1, Joshi \&
Sagar (1977) have given the age 3$\times$10$^{7}$ yr, which is also in good 
agreement.

    To determine the age and distance of the cluster Be 11 with the combination
of optical and Near-IR data, we plot $V$ vs $(V-K)$ and $K$ vs $(J-K)$ in Fig 11
. We have overplotted the theoretical isochrones of log(age) = 8.05 given by 
Schaller et al. (1992). The apparent distance moduli $(m - M)_{V, (V-K)}$ and 
$(m-M)_{K, (J-K)}$ turn out to be 14.6$\pm$0.3  and 12.0$\pm$0.3 mag 
respectively. By using the reddening estimated in the previous section we 
derive a distance of 2.1$\pm$0.3 Kpc for Be 11. Both age and distance 
determination for Be 11 are thus in excellent agreement with our earlier 
estimates.

\begin{figure*}
\hspace{1.0cm}\psfig{file=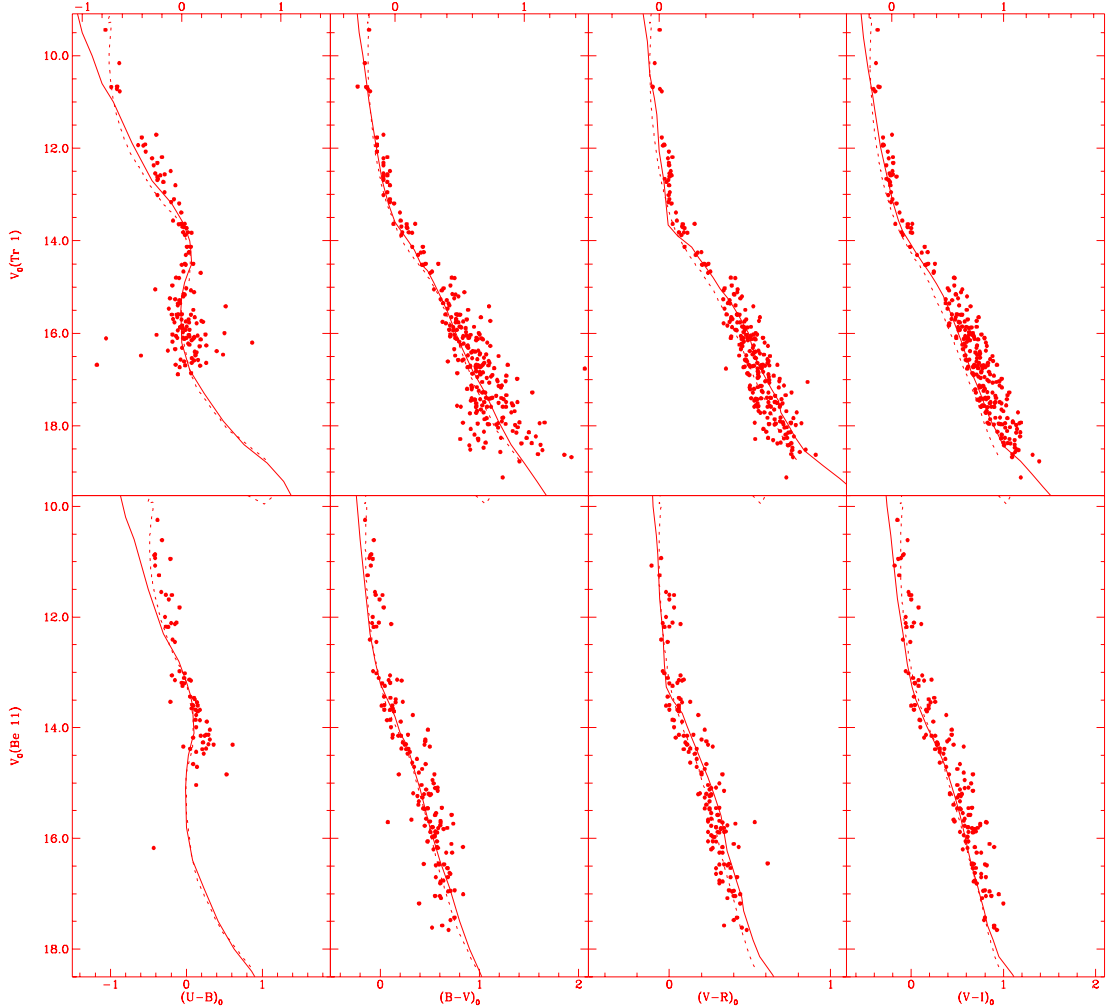,width=15cm,height=15cm}
\caption{The $V_{0}$, $(U-B)_{0}$; $V_{0}$, $(B-V)_{0}$; $V_{0}$, $(V-R)_{0
}$ and $V_{0}$, $(V-I)_{0}$ diagrams for stars of the Tr 1 and Be 11. The 
continuous curves are the ZAMS fitted to the MS. The dotted curves are the 
isochrones for Pop I stars of log(age)=7.6 for Tr 1 and 8.05 for Be 11 fitted 
to the brighter cluster members.}
\end{figure*}

\subsection {Luminosity and Mass function of the clusters}

     The luminosity function (LF), denotes the relative number of stars
in unit magnitude range. The
correction of non-member stars is very much important in the construction of
LF's for star clusters. Two colours, such as $B$ and $V$ or $V$ and $I$, are
required for the non-member identification. It is therefore required to
construct the LF either from a $V$, $(B-V)$ diagram or from $V$, $(V-I)$
diagram or from a similar diagram instead of from a single $B$, $V$ or $I$ or
any other passband. We preferred the
$V$, $(V-I)$ over the other diagrams as it is deepest. The main
disadvantage of using two
passbands for the construction of the LF is that both passbands introduce
incompleteness, whose determination is a difficult process, as described below.

\subsubsection {Determination of photometric completeness}

       The method consists of insertion of randomly selected artificial stars
with known magnitude and position in the original $V$ frame. For the $I$ band 
image
the inserted stars have same geometrical positions but differ in $I$ brightness
according to mean $(V-I)$ colour of the MS stars. Only 15\% of the number
of actually detected stars are inserted at one time, so that the crowding 
characteristics
 of the original data remains almost unchanged. The luminosity
distribution of the artificial stars has been chosen in such a way that more
stars are inserted into the fainter magnitude bins. The frames are re-processed
using the same procedure used for the original frames. The ratio of recovered
to inserted stars in the different magnitude bins gives directly the completeness factor (CF) for that region. Table 12 lists the CF in both cluster and field 
regions of the objects under study.

\begin{figure}
\hspace{1.0cm}\psfig{file=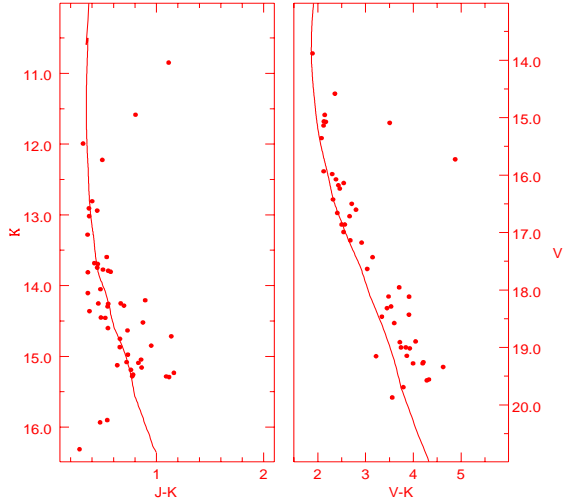 ,width=8.5cm,height=8cm}
\caption{The $K$, $(J-K)$ and $V$, $(V-K)$ diagrams of the sample stars in
the cluster Be 11.}
\end{figure}

  Several authors have discussed the method for determining the CF (cf. Stetson
 1987; Mateo 1988; Sagar \& Richtler 1991; Banks et al. 1995). Sagar \& Richtler
 (1991)  have taken the minimum value of the
completeness factors of the pair to correct
the star counts. They argue that the two frames are not independent and that
the multiplicative assumption of Mateo (1988) could not be justified. Banks et
al. (1995) tested the ability of the techniques given by Mateo (1988) and Sagar
 \& Richtler (1991) on the basis of numerical simulations. They conclude that
product method of Mateo (1988) increasingly over-estimates the incompleteness
correction as the magnitude is increased, and the method suggested by Sagar \&
Richtler (1991) recovered the actual LF better with a mean error of 3\% upto
CF $>$ 0.5. We therefore used the procedure of Sagar \& Richtler (1991) in the
present work.

\begin{table}
 \centering
\caption {Variation of completeness factor (CF) with the MS brightness in
both cluster and field regions.}
\begin{tabular}{|c|c|c|c|c|}
\hline
&\multicolumn{2}{c|}{Tr 1}&\multicolumn{2}{c|}{Be 11}\\\cline{2-5}
$V$ mag range&Cluster&Field&Cluster&Field\\
\hline
13 - 14&0.96&0.99&0.99&0.99\\
14 - 15&0.96&0.99&0.99&0.99\\
15 - 16&0.95&0.99&0.99&0.99\\
16 - 17&0.94&0.98&0.98&0.99\\
17 - 18&0.94&0.98&0.97&0.99\\
18 - 19&0.90&0.98&0.96&0.98\\
19 - 20&0.80&0.98&0.93&0.98\\
\hline
\end{tabular}
\end{table}

\subsubsection{Determination of Mass Function}

           To derive the true LF of the cluster we have to remove the field
star contamination. For this, we use photometric criterion. First, we defined
a blue and red envelope for the MS on $V$, $(V-I)$ CM diagram.
 Same envelope were drawn on the corresponding field region.
Star counts as a function of luminosity were made in both (cluster and field)
regions. In this way we can estimate the number of field stars
present in various magnitude bins of the cluster region. The observed LFs
 of the cluster and field regions were also corrected for data
incompleteness as well as for differences in area. True LF for the cluster was 
obtained by subtracting the observed
LF of field region from the observed LF of cluster region.

\begin{figure}
\psfig{file=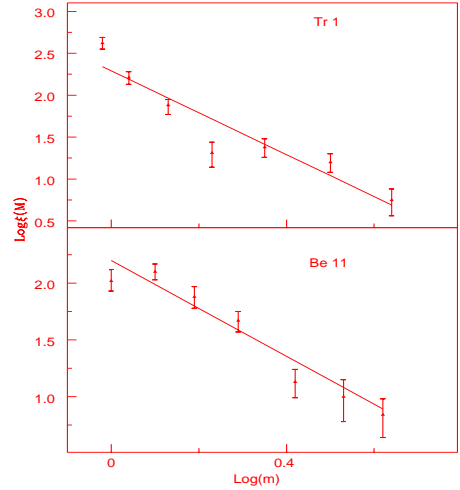 ,width=8.5cm,height=7cm}
\caption{The plot shows the mass functions derived using Schaller et al. 
(1992) isochrones.} 
\end{figure}

    The mass function (MF), which denote the relative number of stars in
unit range of mass centered on mass M. The MF slope has been derived from the 
mass distribution $\xi$($M$). If $dN$ represents the number of stars 
in a mass bin $dM$ with central mass $M$, then the value of slope $x$ is 
determine from the linear relation 

\vspace{0.2cm}
~~~~~~~~~~~~~~~~log$\frac{dN}{dM}$ = $-$(1+$x$)$\times$log($M$)$+$constant

\vspace{0.2cm}
\noindent using the least-squares solution. The Salpeter (1955) value for the 
slope of MF is $x$ = 1.35.\\

        To derive the MF from LF, we need theoretical
evolutionary tracks and accurate knowledge of cluster parameter like reddening
, distance, age etc. Theoretical models by Schaller et al. (1992) were used to
convert the observed LF to the MF. Fig 12 represents the plot of MFs of Tr 1 and
 Be 11.
 The value of the MF slope along with the mass range and error are
given in Table 13, where the quoted errors are errors resulting from the linear
least square fit to the data points. For the cluster Tr 1, the value of $x$
 is in agreement with the value given by Phelps \& Janes (1993). The values of 
$x$ for both clusters are in agreement with the Salpeter value.

The dip is seen in the MF of both the clusters. In the cluster Tr 1 one dip 
is present in the mass range 1.5$<M/M_{\odot}<$1.9. This dip is due to 
the deficiency of stars in the intrinsic $V$ magnitude between 14.0 and 
14.8 (see Sec. 4.5). In this cluster an apparent dip in LF is also found by 
Phelps \& Janes (1993) at 1.4 $M_{\odot}$. One dip is also seen in the MF of Be 
11 in the mass range 2.0$<M/M_{\odot}<$3.0. This dip corresponds to the gap 
noticed in the intrinsic $V$ magnitude between 12.4 and 12.9 (see Sec. 4.5).

\begin{table}
 \centering
\caption{The slope of the mass function derived from LF alongwith relaxation 
time T$_{E}$.}

\begin{tabular}{cccc}
\hline
cluster&Mass range&Mass Function slope &log T$_{E}$\\
&M$_{\odot}$&($x$)&\\
\hline
Tr 1&0.9 - 5.1&1.50$\pm$0.40&7.2\\
Be 11&1.0 - 4.5&1.22$\pm$0.24&7.1\\
\hline
\end{tabular}
\end{table}

\section {\bf Mass segregation}
      To study the effect of mass segregation for the clusters under study,
 we plot in Fig. 13 cumulative radial stellar distribution of stars for
different masses. A careful inspection of Fig 13 shows that both the
clusters have mass segregation effect.\\

   To perform the K-S test among these distribution to see whether they belong
to the same distribution or not, we have divided in
 three mass range 5.0$\le$ M$_{\odot}$$<$2.5, 2.5$\le$ M$_{\odot}$$<$1.0 and
M$_{\odot}$$<$1.0 and 4.5$\le$ M$_{\odot}$$<$2.5, 2.5$\le$ M$_{\odot}$$<$1.5 and
 M$_{\odot}$$<$1.5 mag for Tr 1 and Be 11 respectively. The K-S test indicates
that mass
segregation has occurred at confidence level of 99\% for Tr 1 and 80\% for Be 11.
One would like to know whether existing mass segregation is due to dynamical 
evolution or
imprint of star formation process.\\

   Dynamical evolution is one of the possible cause for mass segregation. At the
 time of formation, the cluster may have a uniform spatial stellar mass
distribution, which may be modified due to dynamical evolution of the cluster 
members. Because of dynamical relaxation, low mass stars in a cluster may 
possess largest random velocities, consequently these will try to occupy a 
large volume than the high mass stars (cf. Mathieu \& Latham 1986, McNamara \& 
Sekiguchi 1986, Mathieu 1985). Thus mass segregation develops in the time 
scale required to exchange energy between stars of different mass by 
scattering. The dynamical relaxation time, T$_{E}$ is the time in which the 
individual stars exchange energies and their velocity distribution approaches 
a maxwellian equilibrium. It is given by\\
\begin{displaymath}
~~~~~~~~~~~~~~~~~~~~T_{E} = \frac {8.9 \times 10^{5} N^{1/2} R_{h}^{3/2}}{ <m>^{1/2}log(0.4N)}
\end{displaymath}

where $N$ is the number of cluster members, $R$$_{h}$ is the radius containing 
half of the cluster mass and $<m>$ is the average mass of the cluster stars (cf.
Spitzer \& Hart 1971). The number of probable MS stars is estimated using the 
CMDs of the clusters after subtracting the contribution due to field stars
 and applying the necessary corrections for the data incompleteness. Due to our
inability to estimate the R$_{h}$ from the present data, we assume that the R$_{
h}$ is equal to half of the cluster radius derived by us. The angular values
are converted to linear values using the cluster distances which are derived 
here. Inclusion of cluster members fainter than the
limiting $V$ magnitude will decrease the value of $<m>$ and increase the value
 of $N$. This will result in higher values of $T_{E}$. Hence the $T_{E}$ values obtained
here may be considered as the lower limit.\\

\begin{figure}
\psfig{file=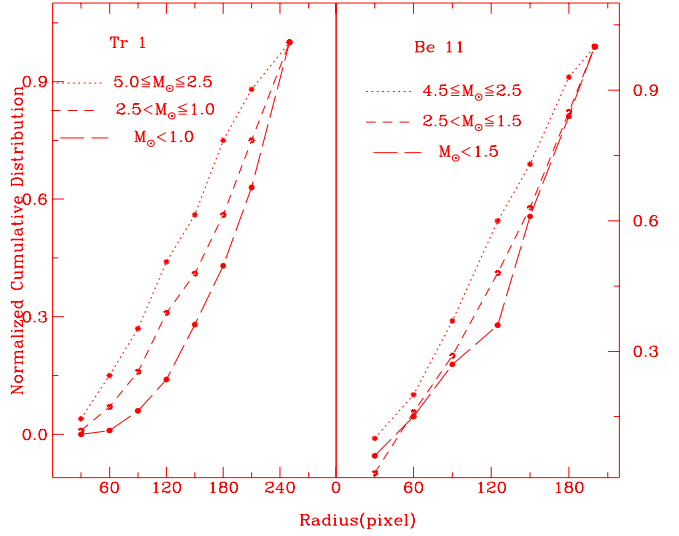 ,width=9.5cm,height=9cm}
\caption{Cumulative radial distribution of stars in different mass ranges for Tr 1 and Be 11.}
\end{figure}

          A comparison of cluster age with its relaxation time indicates that
the relaxation time is smaller than the age of the clusters. Thus
we can conclude that the clusters under study are dynamically relaxed. It may be
 due to the result of dynamical evolution or imprint of star formation processes
 or both.
\section {Conclusions}

       Physical parameters of the young open star cluster Tr 1 and Be 11 have
been derived based on $UBVRI$ CCD photometry. Our study leads to the following
conclusions:\\

   (i) The radial density profiles of the clusters indicate the existence of
clustering. The angular diameter of the cluster Tr 1 and Be 11 are 6$^\prime$.0 and 4$^{\prime}$.8 
respectively. The corresponding linear sizes are 4.6 and 3.0 pc respectively.\\

   (ii) Variable reddening is present within both the clusters Tr 1 and Be 11
with
the mean value of $E(B-V)$ = 0.60$\pm$0.05 mag and 0.95$\pm$0.05 mag 
respectively. The law of interstellar extinction is normal in the direction of 
both the clusters. Combining 2MASS data with optical data, we have studied the extinction law
in the direction of Be 11. Colour-colour diagram yields
 the colour excess $E(J-K)$ = 0.40$\pm$0.20 mag and $E(V-K)$ = 2.20$\pm$0.20 mag respectively.\\

(iii) The distance values are 2.6$\pm$0.10 and 2.2$\pm$0.10 Kpc for the cluster
Tr 1 and Be 11 respectively. The fitting of Schaller et al. (1992) isochrones to the intrinsic 
CM diagrams indicate an age of 40$\pm$10 and 110$\pm$10 Myrs for for Tr 1 and Be 11 respectively.\\

 (iv) The luminosity function of clusters was constructed by subtracting field
star contamination determined from neighbour field's. The luminosity
function was transformed into the mass function using the present cluster
parameters and the theoretical model
given by Schaller et al. (1992). This gives the mass function slope $x$
= $1.50\pm0.40$ and 1.22$\pm0.24$ for Tr 1 and Be 11 respectively. The
values of $x$ for Tr 1 and Be 11 thus agrees with the Salpeter (1955) value.\\

(v) Mass segregation, in the sense that massive stars tend to lie near the
cluster center, is observed in both the cluster Tr 1 and Be 11. The dynamical
relaxation time indicate that both the clusters are dynamically relaxed and
mass segregation may have occurred due to dynamical evolution, or imprint of
star formation or both.\\

(vi) Present data are unable to unambiguously identify the members of
the clusters. For this, precise kinematical observations are required.
\section*{Acknowledgments}

We gratefully acknowledge the useful comments given by the referee J. C. Mermilliod, which improved the paper 
significantly. We are also grateful to Dr. Vijay Mohan for helping in data reduction. Useful discussions 
with Dr. A. K. Pandey is thankfully acknowledged. This study made use of 
2MASS and WEBDA.

\bsp
\label{lastpage}


\begin{thebibliography}{99}
\bibitem{} Alcala$^{\prime}$ J. M., Ferro A. A., 1988, Rev. Mex. Astro. Astrofis 16, 81
\bibitem{} Banks T., Dodd R. J., Sullivan D. J., 1995, MNRAS 274, 1225
\bibitem{}  Burki G., 1975, A\&A 43, 37
\bibitem{} Caldwell A. R. John, Cousins A. W. J., Ahlers C. C., Wamelen P. van, Maritz E. J., 1993, SAAO Circ. No. 15
\bibitem{} Cardelli J. A., Clayton G. C., Mathis J. S., 1989, ApJ 345, 245
\bibitem{} Hawarden T. G., 1971, Observatory 91, 78
\bibitem{} Jackson P. D., Fitzgerald M. P., Moffat A. F. J., 1980, A\&AS 41, 211
\bibitem{} Johnson H. L., Morgan W. W., 1953, ApJ 117, 313
\bibitem{} Joshi U. C., and Sagar R., 1977, Ap\&SS 48, 225
\bibitem{} Kamp L. W., 1974, A\&AS 16, 1
\bibitem{} Kaluzny J., 1992, Acta Astron. 42, 29
\bibitem{} Landolt A. U., 1992, AJ 104, 340
\bibitem{} Mateo M., 1988, ApJ 331, 261
\bibitem{} Mathieu R. D., 1985, Dynamics of star clusters 113, 427 J. Goodman 
and P. Hut (eds.)
\bibitem{} Mathieu R. D., \& Latham D. W., 1986, AJ 92, 1364
\bibitem{} McCuskey S. W. and Houk N., 1964, Astron. J. 69, 412
\bibitem{} McNamara B. J. \& Sekiguchi K., 1986, ApJ 310, 613
\bibitem{} {Mermilliod J. C., 1995, in Information and on - line data in 
Astronomy, Eds E. Egret and M. A. Abrecht, Kulwer Academic Press, p 227.}
\bibitem{} Oja T., 1966, Arkiv Astron. 4, 15
\bibitem{} Moitinho, 2001, A\&A 370, 436
\bibitem{} Nilakshi, Sagar R., Pandey A. K., Mohan V., 2002, A\&A 383, 153
\bibitem {}Phelps R. L. and Janes K., 1993, AJ 106, 1870
\bibitem{} Phelps R. L. and Janes K., 1994, ApJS 90, 31
\bibitem{} Persson S. E., Murphy D. C., Krzeminski W., Roth M., \& Rieke M. J., 
1998, AJ 116, 2475.
\bibitem{} Sagar R., Joshi U. C., 1979, Ap\&SS 66, 3
\bibitem{} Sagar R., 1987, MNRAS 228, 483
\bibitem{} Sagar R., Miakutin V. I. Piskunov A. E., Dluzhnevskaia O. B 1988, MNRAS 234, 831
\bibitem{} Sagar R. and Qian, 1990, ApJ 353, 174
\bibitem{} Sagar R. and Richtler T., 1991, A\&A 250, 324
\bibitem{} Sagar R. 2001 in IAU Symp. 207, Extragalactic Star Clusters, eds 
Eva grebel (in press)
\bibitem{} Salpeter E. E., 1955, ApJ 121, 161
\bibitem{} Schaller G., Schaerer D., Meynet G., Maeder A., 1992, A\&AS 96, 269
\bibitem{} Schmidt - Kaler Th., 1982, In: Landolt/Bornstein, Numerical Data and
Functional Relationship in Science and Technology, New series, Group VI, Vol.
2b, Scaifers K. \& Voigt H. H. (eds.) Springer - Verlog, Berlin, p. 14
\bibitem{} Spitzer L. and Hart M. H., 1971, ApJ 164, 399
\bibitem{} Steppe H., 1974, A\&AS 15, 91
\bibitem{} Stetson P. B., 1987, PASP 99, 191
\bibitem{} Walker A. R., 1985, MNRAS 213, 889
\bibitem{} Walker A. R., 1987, MNRAS 229, 31
\bibitem{} Whittet D. C. B., van Breda I. G., 1980, MNRAS 192, 467
\bibitem{} Yadav R. K. S. and Sagar R., 2001, MNRAS 328, 370
\end{thebibliography}
\end{document}